\documentclass[12pt]{iopart}

\usepackage{amsmath}
\usepackage{iopams}
\usepackage{multicol,float,graphicx,color,epstopdf}

\begin{document}

\maketitle

\title{Lagrangian Formalism in Biology: II. Non-Standard and Null Lagrangians and their Role in Population Dynamics}

\author{D. T. Pham}
\address{Department of Biology, The University of Texas at Arlington, TX 76019, USA}
\ead{npham@mavs.uta.edu}

\author{Z. E. Musielak}
\address{Department of Physics, The University of Texas at Arlington, TX 76019, USA}
\ead{zmusielak@uta.edu}

\begin{abstract}
Non-standard Lagrangians do not display any discernible 
energy-like terms, yet they give the same equations of motion
as standard Lagrangians, which have easily identifiable energy-like 
terms.  A new method to derive non-standard Lagrangians for 
second-order nonlinear differential equations with damping is
developed and the limitations of this method are explored.  It is 
shown that the limitations do not exist only for those nonlinear 
dynamical systems that can be converted into linear ones. The 
obtained results are applied to selected population dynamics 
models for which non-standard Lagrangians and their corresponding 
null Lagrangians and gauge functions are derived, and their roles 
in the population dynamics are~discussed.
\end{abstract}

\section{Introduction}\label{sec1}

In physics equations of motion are derived by using the Lagrangian formalism,
which requires knowledge of a function called Lagrangian. By specifying a 
Lagrangian for a given system, its equation of motion is derived by substituting 
this Lagrangian into the Euler-Lagrange (E--L) equation. There are three different 
families of Lagrangians, namely, standard, non-standard, and null.  

Standard Lagrangians represent the difference between the kinetic and potential 
energy [1], and they play a central role in Classical Mechanics [2-4] as well as 
other areas of physics [5]. Non-standard Lagrangians (NSLs) are different in form 
from standard Lagrangians but also give the same equation of motion as the NSLs.  
The main difference is that NSLs lack terms that clearly discern energy-like forms [6].
There are also null Lagrangians and they identically satisfy the E-L equation and 
can be expressed as the total derivative of any scalar function [7].  Since null 
Lagrangians have no effect on equations of motion, they are not discussed any 
further in this paper.

Different methods to derive standard [8-15] and nonstandard [12-26] Lagrangians 
were developed. The methods were used to obtain both types of Lagrangians for
many different physical systems. In the first attempts to obtain Lagrangians for 
selected biological systems, the Lagrangians were found by guessing their forms 
[27-29]. Then, they were derived using the Method of Jacobi Last Multiplier [30,31].
More recently, the first standard Lagrangians were derived for five population 
dynamics models and their implications for the models were discussed [32].  

This paper aims to extend this work by deriving NSLs for selected population 
dynamics models considered in [32], which are: the Lotka-Volterra [33,34], 
Verhulst [35], Gompertz [36], Host-Parasite [37], and SIR [38] models. The 
models are described by second-order nonlinear and damped ODEs, for which a
new method is required to find NSLs. In this paper, a novel method is developed 
and then used to derive NSLs for all considered population models. The derived 
NSLs are compared to those previously obtained [29,30,32] and used to gain new 
insights into the role they play in the models, and in the symmetries underlying 
them. For the derived NSLs, their corresponding null Lagrangians (NLs) and gauge 
functions are also calculated, and these NLs in the population dynamics 
is discussed.

The paper is organized as follows. Section 2 presents our method to construct 
non-standard Lagrangians and its limitations; in Section 3, the population 
dynamics models are described and our method to derive their non-standard 
Lagrangians for these models are presented; then, non-standard Lagrangians 
for the population dynamics models are given in Section 4; null Lagrangians 
and their gauge functions are derived and discussed in Section 5; the obtained 
results are compared to those previously obtained and discussed in Section 6; 
and our Conclusions are given in Section 7.

\section{Lagrangian formalism}\label{sec2}

\subsection{Action and non-standard Lagrangians}\label{sec2.1}

The functional $\mathcal{S}[x(t)]$ is called action and is defined by 
an integral over a scalar function $L$ that depends on the differentiable 
function $x(t)$ that describes the time evolution of any dynamical system,
and on the time derivative $\dot x(t) = dx/dt$.  The function $L(\dot{x},
x)$ is called the Lagrangian function or simply {\it Lagrangian}, and it 
is a map from the tangent bundle $T\mathcal{Q}$ to the real line 
$\mathbf{R}$, or $\mathcal{L}:T\mathcal{Q} \rightarrow \mathbf{R}$, 
with $\mathcal{Q}$ being a configuration manifold [3].   In general, 
the Lagrangian may also depend explicitly on $t$, which can be written 
as $L(\dot{x},x,t)$ and requires $\mathcal{L}:T \mathcal{Q}\times 
\mathbf{R}\rightarrow \mathbf{R}$. 

According to the principle of least action, or Hamilton's principle [2-6], 
the action $\mathcal{S}[x(t)]$ must obey the following requirement 
$\delta\mathcal{S}=0$, which guarantees that the action is stationary, or 
has either a minimum, maximum, or saddle point.  The necessary condition 
that $\delta\mathcal{S} = 0$, is known as the E--L equation, whose operator 
$\hat {EL}$ acts on the Lagrangian, $\hat {EL} [ \mathcal{L} (\dot x, x, t) ] 
= 0$ and gives a second-order ODE that becomes an equation of motion for 
a dynamical system with $\mathcal{L} (\dot x, x, t)$. The process of deriving 
the equation of motion is called the Lagrangian formalism [2-4], and it has 
been extensively used in physics to derive its fundamental classical and 
quantum equations [5].

The Lagrangian formalism is valid for both standard and non-standard 
Lagrangians but only the latter are considered in this paper.  There are 
different forms of NSLs [12-22] but only two of them are considered in
this paper.  The first considered form of NSL is 

\begin{equation}
L_{ns1} (\dot x,x,t)=\frac{1}{A(t)\dot x+B(t)x+C(t)}\ ,
\label{eq1a}
\end{equation}
and it is applied to the equation of motion given by 
\begin{equation}
\ddot x+a(t)\dot x^2+b(t)\dot x+c(t)x= 0\ ,
\label{eq1b}
\end{equation}
where the coefficients $a(t)$, $b(t)$ and $c(t)$ are at least twice differentiable 
functions of the independent variable $t$, and the functions $A(t)$, $B(t)$ 
and $C(t)$ are expressed in terms of these coefficients by substituting 
$L_{ns1} (\dot x,x,t)$ into the E--L equation.

The second form of the considered NSL is 
\begin{equation}
L_{ns2}(\dot x,x)=\frac{1}{f(x)\dot x+g(x)x+h(x)}\ ,
\label{eq2a}
\end{equation}
and it is used for the following equation of motion 
\begin{equation}
\ddot x+\alpha (x)\dot x^2+\beta(x)\dot x+\gamma(x)x=0\ ,
\label{eq2b}
\end{equation}
where the coefficients $\alpha(x)$, $\beta(x)$ and $\gamma(x)$ 
are at least twice differentiable functions of the dependent variable 
$x(t)$, and the functions $f(x)$, $g(x)$ and $h(x)$ are determined 
by substituting $L_{ns2} (\dot x,x)$ into the E--L equation, which 
allows expressing these functions in terms of the coefficients.   

The NSL $L_{ns1} (\dot x,x,t)$ was extensively studied and applied 
to several dynamical systems [12,13,20].  However, studies of
$L_{ns2}(\dot x,x)$ were limited to only some special cases [20].  
Therefore, one of the objectives of this paper is to perform a detailed 
analysis of $L_{ns2}(\dot x,x)$ and its applicability to the population 
dynamics models; so far, only other forms of NSLs been applied to
the models [27-31].  Moreover, in a recent work [32], the standard 
Lagrangians were constructed for the same sample of the population

\subsection{Limits on construction of non-standard Lagrangians}\label{sec2.2}

To obtain a NSL means to determine the unknown functions in 
Eqs (\ref{eq1a}) and (\ref{eq2a}), which requires that an equation 
of motion is given.  As shown in [32], the equations of motion 
for the population dynamics models can be written in the following form 
\begin{equation}
\ddot x+\alpha (x)\dot x^2+\beta(x)\dot x+\gamma(x)x= F(x)\ ,
\label{eq3}
\end{equation}
which generalizes Eq. (\ref{eq2b}) by taking into account a driving 
force $F(x)$.  For different models, the coefficients $\alpha(x)$, 
$\beta(x)$ and $\gamma(x)$ are given, and $F(x)$ is also known.
Since Eq. (\ref{eq3}) does not depend explicitly on time, the 
Lagrangian that can be used for this equation is $L_{ns2}(\dot x,x)$
given by Eq. (\ref{eq2a}).  From now on, we shall take $L_{ns2}
(\dot x,x)\equiv L_{ns}(\dot x,x)$ and use the latter throughout
this paper.   

It is seen that Eq. (\ref{eq3}) has both linear and quadratic damping 
terms, that can become nonlinear depending on the form of 
$\gamma(x)x$, and that it is driven.  In a previous study [32], the 
standard Lagrangian for Eq. (\ref{eq3}) was found in case $ F(x)$ 
= const.  To construct NSL for Eq. (\ref{eq3}), we take $L_{ns}(\dot x,x)$ 
given by Eq. (\ref{eq2a}) and determine the functions $f(x)$, $g(x)$ 
and $h(x)$ in terms of the known coefficients $\alpha(x)$, $\beta(x)$ 
and $\gamma(x)$.  Our approach generalizes the previous work [18],
and it allows us to investigate the applicability of $L_{ns}(\dot x,x)$ 
to Eq. (\ref{eq3}) and its limitations.\\ \\ \\
{\bf Proposition 1:}  Let a general non-standard Lagrangian be given by
\begin{equation}
L_{ns}(\dot x,x)=\frac{1}{f(x)\dot x + G(x)}\ ,
\label{eq4}
\end{equation}
where $G(x) = g(x)x+h(x)$.  Then, $L_{ns}(\dot x,x)$ becomes the 
NSL for Eq. (\ref{eq3}) if, and only if, 
\begin{equation}
f (x) = e^{\int^x \alpha (\xi) d \xi} \equiv e^{I_{\alpha} (x)}\ ,
\label{eq5}
\end{equation}
and
\begin{equation}
G (x) = 3 \left [ \frac{\gamma (x) x -  F(x)}{\beta (x)} \right ]
e^{I_{\alpha} (x)}\ ,
\label{eq6}
\end{equation}
and either one of the following conditions 
\begin{equation}
\beta (x) = \frac{9}{2} e^{- I_{\alpha} (x)} \frac{d}{dx} \left [ 
\left ( \frac{\gamma (x) x -  F(x)}{\beta (x)} \right )
e^{I_{\alpha} (x)} \right ]\ ,
\label{eq7}
\end{equation}
and
\begin{equation}
\gamma (x) x - F(x) = \frac{2}{9} \beta (x) e^{- I_{\alpha} (x)} 
\int^x \beta (\xi) e^{I_{\alpha} (\xi)} d\xi\ ,
\label{eq8}
\end{equation}
is satisfied.\\ \\
{\bf Proof:} Substituting Eq. (\ref{eq3}) into the E--L equation, 
we obtain 
\begin{equation}
\ddot x + \left [ \frac{f^{\prime (x)}}{f(x)} \right ] \dot x^2 + 
\left [ \frac{3}{2f (x)} \frac{G^{\prime} (x)}{f(x)} \right ] \dot x
+ \left [ \frac{1}{2 x} \frac{G^{\prime} (x)}{f(x)} \frac{G (x)} {f (x)} 
\right ] x = 0\ .
\label{eq9}
\end{equation}
By comparing this equation to Eq. (\ref{eq3}), we find the following 
relationships
\begin{equation}
\alpha (x) = \frac{f^{\prime (x)}}{f(x)}\ , 
\label{eq10a}
\end{equation}
\begin{equation}
\beta (x) = \frac{3}{2f (x)} \frac{G^{\prime} (x)}{f(x)}\ ,
\label{eq10b}
\end{equation}
and 
\begin{equation}
\gamma (x) = \frac{1}{2 x} \frac{G^{\prime} (x)}{f(x)} 
\frac{G (x)}{f(x)} + \frac{F (x)}{x}\ .
\label{eq10c}
\end{equation}
Using Eq. (\ref{eq10a}), we obtain $f(x)$ given by Eq. (\ref{eq5}).
However, by combining Eqs (\ref{eq10b}) and (\ref{eq10c}), and 
using Eq. (\ref{eq10a}), we find $G(x)$ given by Eq. (\ref{eq6}).
Then, the conditions expressed by Eqs (\ref{eq7}) and (\ref{eq8})
are easy to derive from Eqs (\ref{eq10a}), (\ref{eq10b}) and 
(\ref{eq10c}).  Moreover, substituting Eq. (\ref{eq8}) into Eq. 
(\ref{eq7}) shows that both conditions are equivalent, thus, it is 
sufficient to use only one of them. This concludes the proof.\\ \\
{\bf Corollary 1:} The coefficients $\beta (x)$ and $\gamma (x)$ 
mutually depend on each other through the function $G(x)$ and, 
in addition, they also depend on $\alpha (x)$ through the function
$f(x)$.\\ \\
{\bf Corollary 2:} The conditions given by Eqs (\ref{eq7}) and 
(\ref{eq8}) are equivalent, which means that if one of them is not 
satisfied then $L_{ns}(\dot x,x)$, with $f(x)$ and $G(x)$ (see 
Eqs \ref{eq5} and \ref{eq6}), cannot be considered to be the NSL
for Eq. (\ref{eq3}).\\ \\
{\bf Corollary 3:} Only $\alpha (x)$ is required to uniquely 
determine the function $f(x)$.

\subsection{From nonlinear to linear equations}

The main result of Proposition 1 is that any NSL of the form of Eq. 
(\ref{eq4}) can only be constructed if one of the conditions given by 
Eqs (\ref{eq7}) and (\ref{eq8}) is satisfied.  We now explore other 
consequences of the condition given by Eq. (\ref{eq8}) in the 
following proposition.\\ 
{\bf Proposition 2:}  Let the equation of motion be  
\begin{equation}
\ddot x+\alpha (x)\dot x^2+\beta(x)\dot x+H(x) = 0\ ,
\label{eq11}
\end{equation}
where $H (x) = \gamma (x) x  - F (x)$, and let  
\begin{equation}
z (x) = \int^x \beta (\xi) e^{I_{\alpha} (\xi)} d\xi\ ,
\label{eq12}
\end{equation}
be a new dependent variable that can be expressed in terms of 
a new variable $\eta$ that is related to the original variable $x$ 
by $d \eta = \beta (x) dx$.  Then, Eq. (\ref{eq11}) becomes  
\begin{equation}
z^{\prime \prime} + z^{\prime} + z = 0\ ,
\label{eq13}
\end{equation}
with $z^{\prime} = dz / d\eta$ and $z^{\prime \prime} = d^2 z / 
d\eta^2$, if, and only if,  
\begin{equation}
H (x) = \frac{2}{9} \beta (x) e^{- I_{\alpha} (x)} 
\int^x \beta (\xi) e^{I_{\alpha} (\xi)} d\xi\ .
\label{eq14}
\end{equation}
{\bf Proof:} Using Eq. (\ref{eq12}), we find 
\begin{equation}
z^{\prime} = \dot x e^{I_{\alpha} (x)}\ ,
\label{eq15a}
\end{equation}
and 
\begin{equation}
z^{\prime \prime} = \frac{1}{\beta(x)} \left [ \ddot x + \alpha (x) 
\dot {x}^2 \right ] e^{I_{\alpha} (x)}\ .
\label{eq15b}
\end{equation}
Substituting Eqs (\ref{eq15a}), (\ref{eq15b}) and (\ref{eq12}) into
Eq. (\ref{eq13}), we obtain the original equation of motion (see Eq. 
\ref{eq11}) with the function $H(x)$ given by Eq. (\ref{eq14}). 
This concludes the proof.\\ \\
{\bf Corollary 4:} Equation (\ref{eq13}) is the Sturm-Liouville equation 
[43] whose solutions are well-known and given as $z (\eta) = c_1 [ \sin
(\sqrt{3} \eta / 2) + c_2 \cos (\sqrt{3} \eta / 2) ] e^{-\eta /2}$.\\ \\
{\bf Corollary 5:} The non-standard Lagrangian given by Eq. (\ref{eq4})
can be constructed without any limits for those nonlinear ODEs that can 
be converted into linear ODEs, which requires that the condition given 
by Eq. (\ref{eq8}) is satisfied.\\ 

\subsection{Non-standard Lagrangians without limits}

The results of Proposition 2 and Corollaries 4 and 5 demonstrate that 
the existence of NSL of the form of Eq. (\ref{eq4}) is determined by 
the conditions given by Eqs (\ref{eq7}) and (\ref{eq8}), which must 
be satisfied in order for the NSL to exist.  According to Corollaries 1-3, 
the only two terms in the equation of motion given by Eq. (\ref{eq3}) 
that can be uniquely derived without any limits by the NSL are the terms 
with $\ddot x (t)$ and $\dot x^2 (t)$.  The coefficients in terms with 
$\dot x(t)$ and $x (t)$ are mutually related, thus, they are dependent 
on each other as shown by either Eq. (\ref{eq7}) or Eq. (\ref{eq8}).

Because of these limitations on the construction of NSLs for Eq. (\ref{eq3}),
let us now propose another method by extending the previous work [32].
The basic idea of this work is to write Eq. (\ref{eq3}) in the following form 
\begin{equation}
\ddot x+\alpha (x)\dot x^2 = {\mathbb F}_{dis} (\dot x, x)\ ,
\label{eq16}
\end{equation}
where 
\begin{equation}
{\mathbb F}_{dis} (\dot x, x) = F(x) - \beta(x)\dot x - \gamma(x)x\ ,
\label{e17}
\end{equation}
becomes a dissipative force because of its dependence on $\dot x (t)$. 
In this case, the E--L equation [2,3,32] can be written as 
\begin{equation}
{d\over dt}\biggr({\partial L\over\partial\dot x}\biggr)-{\partial L
\over\partial x}={\mathbb F}_{dis} (\dot x,x)e^{2I_{\alpha}(x)}\ ,
\label{eq18}
\end{equation}
where the force on the RHS of this equation is known at the Rayleigh 
force [2,3]

Based on the results presented in Section 2.2, the NSL for Eq. (\ref{eq11})
is 
\begin{equation}
L_{ns}(\dot x,x)=\frac{1}{\dot x e^{I_{\alpha} (x)} + C_o}\ ,
\label{eq19}
\end{equation}
where the constant $C_o$ replaces the function $G (x)$ in Eq. (\ref{eq4});
the constant can have any real value, and it is not required to determine 
its value.  This non-standard Lagrangian has no restrictions or limitations 
and it exists for any differentiable $\alpha (x)$ regardless of the forms of 
$\beta (x)$ and $\gamma (x)$.  Therefore, $L_{ns}(\dot x,x)$ will be 
used to find NSLs for the population dynamics models (see Section 3).  
Moreover, the validity of one of the conditions given by Eqs (\ref{eq7}) 
and (\ref{eq8}) must be verified for all the models to determine whether 
any of these models allows for the NSL given by Eq. (\ref{eq4}).  

\section{Population dynamics models and methods}

\subsection{Selected models} 

We consider the following population dynamics models: the Lotka-Volterra 
[33,34], Verhulst [35], Gompertz [36] and Host-Parasite [37] models that 
describe two interacting (preys and predators) species.  However, the SIR 
model [38] describes the spread of a disease in a given population.
%

%
%%%%%%%%%%%%        Table 1       %%%%%%%%%%%%%%%%%
\begin{table}
\caption{Biological models and their equations of motion.\label{tab1}}
\centering
\begin{tabular}{llccc} 
\hline\hline   
{\bf Population models}	&{\bf Equations of Motion}\\ 
\hline
Lotka-Volterra Model&$\dot{w_1}=w_1(a\:+\:bw_2\:)$\\
                    &$\dot{w_ 2}=w_2(A+Bw_1)$\\ \hline
Verhulst Model      &$\dot{w}_1=w_1(A+\:Bw_1\:+f_1 w_2)$\\
                    &$\dot{w}_2=w_2(\:a+\:\:bw_2\:\:+f_2 w_1)$\\ \hline
Gompertz Model      &$\dot{w}_1=w_1{(}A\log{\bigr(}{w_1\over m_1}{\bigr)}+B{w_2}{)}$\\
                    &$\dot{w}_2=w_2{(}\:a\log{\bigr(}{w_2\over m_2}{\bigr)}+\:b{w_1})$\\ \hline
Host-Parasite Model &$\dot{w}_1=w_1(a\:-\:\:bw_2)$\\ 
			    &$\dot{w}_2=w_2(A-B{w_2\over w_1})$ \\ \hline
SIR Model           &$\dot w_1=-b w_1w_2$\\
			      &$\dot w_2=\:\:\:b w_1w_2-aw_2$\\
\hline  
\end{tabular}
\end{table}
%%%%%%%%%%%%%%%%%%%%%%%%%%%%%%%%%%%%%
%
\bigskip
The variables $w_1 (t)$ and $w_2 (t)$ in the first four models of Table 1 
represent prey and predators of the interacting species, respectively. 
Moreover, the time derivatives $\dot w_1 (t)$ and $\dot w_2 (t)$ describe
changes of these species in time.  The interaction between the species in 
each model is given by the coefficients $a$, $A$, $b$, $B$ $f_1$, $f_2$, 
$m_1$ and $m_2$, which are real and constant.  However, in the SIR 
model, the variables $w_1(t)$ and $w_2(t)$ describe susceptible and 
infectious populations, with $a$ and $b$ being the recovery and infection 
rates, respectively. 

Among the models given in Table 1, the Lotka-Volterra, Verhulst and 
Gompertz models are symmetric, and the remaining two models are 
asymmetric, where being symmetric means that the dependent variables 
can be replaced by each other, and the same can be done with the 
coefficients. Obviously, this cannot be done for the asymmetric models.

\subsection{Methods to construct non-standard Lagrangians}

According to Table 1, each model is described by a coupled first-order
nonlinear ODEs that can be cast into a second-order nonlinear ODE of 
the following form
\begin{equation}
\ddot w_{i}+\alpha_{i}(w_{i})\dot w_{i}^2+\beta_{i}(w_{i})\dot w_{i} 
+ \gamma_{i}(w_{i})w_{i}=F_{i}(w_i)\ ,
\label{eq20}
\end{equation}

where $i=1$ and $2$.  This equation is of the same form as that given 
by Eq. (\ref{eq3}).  The non-standard Lagrangian for this equation is 
\begin{equation}
L_{ns}(\dot w_i,w_i)=\frac{1}{f(w_i)\dot w_i + G(w_i)}\ ,
\label{eq21}
\end{equation}
with the functions $f(w_i)$ and $G(w_i)$ being given by Eqs (\ref{eq5})
and (\ref{eq6}), respectively.  Thus, for each considered model and for 
each variable in this model, we determine these functions and obtain 
$L_{ns}(\dot w_i,w_i)$.

Then, we verify the validity of one of the conditions given by Eqs (\ref{eq7}) 
and (\ref{eq8}).  Since these conditions are equivalent, we take only one of 
them and select the condition on $\gamma (w_i)$ (see Eq. \ref{eq8}).  The 
explicit form of this condition used in our calculations is 
\begin{equation}
\gamma (w_i) w_i - F_i (w_i) = \frac{2}{9} \beta (w_i) e^{- I_{\alpha} (w_i)} 
\int^{w_i} \beta (\xi) e^{I_{\alpha} (\xi)} d\xi\ ,
\label{eq22}
\end{equation}
This condition is used to verify whether the derived NSL (see Eq. \ref{eq4})
can be constructed for Eq (\ref{eq3}) or not (see Proposition 1).  Moreover, 
the validity of this condition determines uniquely that the equation of motion 
can be converted into a linear second-order ODE, whose solutions are easy to
find (see Proposition 2).

As shown in Section 2.3, another method to construct NSLs is to cast Eq. 
(\ref{eq20}) into the form 
\begin{equation}
\ddot w_{i}+\alpha_{i}(w_{i})\dot w_{i}^2 ={\mathbb F}_{dis, i}(\dot w_{i}, 
w_{i})\ ,
\label{eq23}
\end{equation}
where 
\begin{equation}
{\mathbb F}_{dis,i}(\dot w_{i}, w_{i}) = F_i (w_i) - \beta_{i}(w_{i})\dot w_{i} 
- \gamma_{i}(w_{i})w_{i}\ .
\label{eq24}
\end{equation}
Then, according to Eq. (\ref{eq14}), the NSL for Eq. (\ref{eq23}) is given by
\begin{equation}
L_{ns,i}(\dot w_i,w_i)=\frac{1}{\dot w_i e^{I_{\alpha} (w_i)} + C_o}\ .
\label{eq25}
\end{equation}
This non-standard Lagrangian is not constrained by any conditions and
it can be derived for all considered population dynamics models.  If this 
NSL is substituted into the following EL equation
\begin{equation}
{d\over dt}\biggr({\partial L_{ns,i}\over\partial\dot w_{i}}\biggr)-{\partial 
L_{ns,i}\over\partial w_{i}}={\mathbb F}_{dis,i} (\dot w_{i}, w_{i}) 
e^{2I_{\alpha}(w_i)}\ ,
\label{eq26}
\end{equation}
then, the equation of motion for the considered model is obtained. 
In the following, this method is used to construct the NSLs for all population
dynamics models considered in this paper. 

\section{Models and their non-standard Lagrangians}

\subsection{\bf Lotka-Volterra Model}

The model is symmetric and it is represented mathematically by a system of 
coupled nonlinear first-order ODEs given in Table 1.  We cast the first-order 
ODEs into the second-order ODEs of the form given by Eq. (\ref{eq23}), and
obtain
\begin{equation}
\ddot w_1 - {1\over w_1} \dot w_1^2 = {\mathbb F}_{dis, 1}(\dot w_{1}, 
w_{1})\ ,
\label{eq27a}
\end{equation}  
where 
\begin{equation}
{\mathbb F}_{dis, 1}(\dot w_{1}, w_{1}) =(Bw_1+A) \dot w_1 + 
a (Bw_1+A) w_1\ ,
\label{eq27b}
\end{equation}  
and 
\begin{equation}
\ddot w_2 - {1\over w_2} \dot w_2^2 = {\mathbb F}_{dis, 2}(\dot w_{2}, 
w_{2})\ ,
\label{eq27c}
\end{equation}
with
\begin{equation}
{\mathbb F}_{dis, 2}(\dot w_{2}, w_{2}) = (b w_2+a) \dot w_2 + 
A (bw_2+a) w_2\ .
\label{eq27d}
\end{equation}  
Using Eq. (\ref{eq5}), the factors $e^{I_{\alpha} (w_i)}$ for both models 
can be calculated, and the obtained results are substituted into Eq. (\ref{eq25}) 
to give
\begin{equation}
L_{ns,1}(\dot w_1,w_1)=\frac{1}{\dot w_1 w_1^{-1} + C_o}\ .
\label{eq28a}
\end{equation}
and
\begin{equation}
L_{ns,2}(\dot w_2,w_2)=\frac{1}{\dot w_2 w_2^{-1} + C_o}\ ,
\label{eq28b}
\end{equation}
which are the NSLs for the Lotka-Volterra model.  It is easy to verify that 
by substituting them into the E--L equation given by Eq. (\ref{eq26}), 
the evolution equations describing the Lotka-Volterra model are obtained 
(see Eqs \ref{eq27a} and \ref{eq27c}). 

\subsection{\bf Verhulst Model}

The system of coupled nonlinear ODEs given in Table 1 shows 
that the model is symmetric. The second-order equations for 
the dynamical variables of this model are:
\begin{equation}
\ddot w_1 - (1 + b) {1\over w_1} \dot w_1^2 = {\mathbb F}_{dis, 1}(\dot w_{1}, 
w_{1})\ ,
\label{eq29a}
\end{equation}  
where 
\[
{\mathbb F}_{dis, 1}(\dot w_{1}, w_{1}) = -\dot w_1(2b-1)Bw_1 - 
f_2w_1^2+(2Ab-a)+(f_2-b)B
\]
\begin{equation}
\hskip0.75in + [(f_2-b)Bw_1^2 + (Af_2-2Ab-a)w_1 + A(a-Ab)] w_1\ ,
\label{eq29b}
\end{equation}  
and 
\begin{equation}
\ddot w_2 - (1 + B) {1\over w_2} \dot w_2^2 = {\mathbb F}_{dis, 2}(\dot w_{2}, 
w_{2})\ ,
\label{eq29c}
\end{equation}
with
\[
{\mathbb F}_{dis, 2}(\dot w_{2}, w_{2}) = -\dot w_2(2B-1)bw_2 - 
f_1w_2^2 + (2aB-A) + (f_1-B)b 
\]
\begin{equation}
\hskip0.75in  + [(f_1-B) bw_2^2 + (af_1 - 2aB - A)w_2 + a(A - aB) ] w_2\ .
\label{eq29d}
\end{equation}  
The non-standard Lagrangians for the evolution equations describing this model
are:
\begin{equation}
L_{ns,1}(\dot w_1,w_1)=\frac{1}{\dot w_1 w_1 ^ {-(1+b)} + C_0}\ .
\label{eq30a}
\end{equation}
and
\begin{equation}
L_{ns,2}(\dot w_2,w_2)=\frac{1}{\dot w_2 w_2 ^ {-(1+B)} + C_o}\ ,
\label{eq30b}
\end{equation}

The derived NSLs give the original evolution equations for the model 
(see Eqs \ref{eq29a} and \ref{eq29c}) after they are substituted into 
the E--L equation given by Eq. (\ref{eq26}).  

\subsection{\bf Gompertz Model}
The mathematical representation of this model given by the coupled and nonlinear 
ODEs in Table 1 show that the model is symmetric.

The equation describing the time evolution of each model variable is given as 
\begin{equation}
\ddot w_1 - {1\over w_1} \dot w_1^2 = {\mathbb F}_{dis, 1}(\dot w_{1}, 
w_{1})\ ,
\label{eq31a}
\end{equation}  
where 
\[
{\mathbb F}_{dis, 1}(\dot w_{1}, w_{1}) = Am_1 + bw_1 + g_1(\dot w_1,w_1) 
\dot w_1 - g_1(\dot w_1,w_1) Aw_1 
\]
\begin{equation}
\hskip0.75in + A\log({w_1\over m_1})\ w^2_1\ ,
\label{eq31b}
\end{equation}  
and 
\begin{equation}
\ddot w_2 - {1\over w_2} \dot w_2^2 = {\mathbb F}_{dis, 2}(\dot w_{2}, 
w_{2})\ ,
\label{eq31c}
\end{equation}
with
\[
{\mathbb F}_{dis, 2}(\dot w_{2}, w_{2}) = am_2 + Bw_2 + g_2
(\dot w_2,w_2) \dot w_2-g_2(\dot w_2,w_2)aw_2 + 
\]
\begin{equation}
\hskip0.75in a\log({w_2\over m_2})\ w^2_2 \ .
\label{eq31d}
\end{equation}  
\begin{equation}
L_{ns,1}(\dot w_1,w_1)=\frac{1}{\dot w_1 w_1^{-1} + C_o}\ .
\label{eq32a}
\end{equation}
and
\begin{equation}
L_{ns,2}(\dot w_2,w_2)=\frac{1}{\dot w_2 w_2^{-1} + C_o}\ ,
\label{eq32b}
\end{equation}

To obtain the original evolution equations given by Eqs \ref{eq31a} 
and \ref{eq31c}) for this model, it is necessary to substitute the 
derived NSLs into the E--L equation (see Eq. \ref{eq26}).  

\subsection{\bf Host-Parasite Model}

This model describes the interaction between a host and its parasite.
The model takes into account the non-linear effects of the host population 
size on the growth rate of the parasite population [22].  The system 
of coupled nonlinear ODEs (see Table 1) is asymmetric in the dependent
variables $w_1$ and $w_2$.  The time evolution equations for these 
variables are:
\begin{equation}
\ddot w_1 - {1\over w_1}\Bigr(1+{B\over bw_1}\Bigr) \dot w_1^2 = 
{\mathbb F}_{dis, 1}(\dot w_{1}, 
w_{1})\ ,
\label{eq33a}
\end{equation}  
where 
\begin{equation}
{\mathbb F}_{dis, 1}(\dot w_{1}, w_{1}) = B{a^2\over b} +
\Bigr(A-{2aB\over bw_1}\Bigr)\dot w_1 + aAw_1  \ ,
\label{eq33b}
\end{equation}  
and
\begin{equation}
\ddot w_2 - {2\over w_2} \dot w_2^2 = {\mathbb F}_{dis, 2}(\dot w_{2}, 
w_{2})\ ,
\label{eq33c}
\end{equation}
with
\begin{equation}
{\mathbb F}_{dis, 2}(\dot w_{2}, w_{2}) =
(bw_2-a-A)\dot w_2 + A(bw_2-a) w_2 \ .
\label{eq33d}
\end{equation}  
\begin{equation}
L_{ns,1}(\dot w_1,w_1)=\frac{1}{\dot w_1  w_1^{-1}e^{B \over w_1} + C_o}\ .
\label{eq34a}
\end{equation}
and
\begin{equation}
L_{ns,2}(\dot w_2,w_2)=\frac{1}{\dot w_2 w_2^{-2} + C_o}\ ,
\label{eq34b}
\end{equation}

Then, the original evolution equations for this model (see Eqs \ref{eq33a} 
and \ref{eq33c}) is obtained by substituting the derived NSLs into the E--L 
equation (see Eq. \ref{eq26}).  

\subsection{\bf SIR Model}
The equation describing the time evolution of each model variable is given as
\begin{equation}
\ddot w_1 - {1\over w_1} \dot w_1^2 = {\mathbb F}_{dis, 1}(\dot w_{1}, 
w_{1})\ ,
\label{eq35a}
\end{equation}  
where 
\begin{equation}
{\mathbb F}_{dis, 1}(\dot w_{1}, w_{1}) = (bw_1 - a )\dot w_1 \ ,
\label{eq35b}
\end{equation}  
and 
\begin{equation}
\ddot w_2 - {1\over w_2} \dot w_2^2  = {\mathbb F}_{dis, 2}(\dot w_{2}, 
w_{2})\ ,
\label{eq35c}
\end{equation}
with
\begin{equation}
{\mathbb F}_{dis, 2}(\dot w_{2}, w_{2}) = -bw_2 \dot w_2 + abw_2^2\ .
\label{eq35d}
\end{equation}  
Using Eq. (\ref{eq5}), the factors $e^{I_{\alpha} (w_i)}$ for both models 
can be calculated, and the obtained results are substituted into Eq. (\ref{eq25}) 
to give
\begin{equation}
L_{ns,1}(\dot w_1,w_1)=\frac{1}{\dot w_1 w_1^{-1} + C_o}\ .
\label{eq36a}
\end{equation}
and
\begin{equation}
L_{ns,2}(\dot w_2,w_2)=\frac{1}{\dot w_2 w_2^{-1} + C_o}\ ,
\label{eq36b}
\end{equation}

The SIR model is asymmetric and the derived NSLs give the original 
evolution equations for this model (see Eqs \ref{eq35a} and \ref{eq35c})
after the NSLs are substituted into the E--L equation given by Eq. 
(\ref{eq26}).  

\section{Null Lagrangians for the population models}

In our previous work [32], we showed how to construct standard Lagrangians
for the population dynamics models.  Moreover, in this paper, we constructed 
the NSLs for the same models. However, there is another family of Lagrangians 
called null Lagrangians (NLs), which make the E--L equation identically zero
and are given as the total derivative of a scalar function [39]; the latter is 
called here a gauge function.  The NLs were extensively studied in mathematics
(e.g., [39-44]) and recently in physics [45-49], but to the best of our knowledge,
NSLs have not yet been introduced to biology and, specifically, to its population
dynamics.  Therefore, in the following, we present the first applications of NLs to 
biology and its population dynamics.
 
Recent studies of null Lagrangians demonstrated that there is a different 
condition that is obeyed by NLs and that this condition plays the same role
for NLs as the E--L equation plays for standard and non-standard Lagrangians
[48,49].  The condition can be written as 
\begin{equation}\
     \frac{d L_{null,i}(\dot w_i,w_i)}{dt} = 
     \frac{\partial L_{null,i}}{\partial t} + \dot w_i 
     \frac{\partial L_{null,i}}{\partial w_i} + \ddot w_i
     \frac{\partial L_{null,i}}{\partial \dot w_i} = 0\ ,
\label{eq37}
\end{equation}
and it shows that the substitution of any NL into Eq. (\ref{eq37}) results in
an equation of motion.  However, the resulting equations of motion may be 
limited because their coefficients are required to obey relationships that are 
different for different equations of motion [48,49].  The previous work also 
demonstrated that an inverse of any null Lagrangian generates a non-standard 
Lagrangian, whose substitution into the E--L equation gives a new equation 
of motion.  However, the reverse is not always true, which means that not 
all NSLs have their corresponding NLs because of the so-called null condition 
that must be satisfied [48,49].

For any non-standard Lagrangian of the form
\begin{equation}
L_{ns,i}(\dot{w_i},w_i,t) = \frac{1}{B_i (w_i,t) \dot{w_i} + C_i (w_i,t) w_i}\ ,
\label{eq38}
\end{equation}
the null condition [48,49] is 
\begin{equation}
\left ( \frac{\partial B_i (w_i,t)}{\partial t} \right ) = \frac{\partial [w_i C_i (w_i,t)]}
{\partial w_i}\ .
\label{eq39}
\end{equation}
Comparison of Eq. (\ref{eq38}) to Eq. (\ref{eq25}) shows that $B_i (w_i) =
exp[ I_{\alpha}(w_i)]$ and $w_i C_i (w_i,t) = C_o$, which means that 
the denominator of Eq. (\ref{eq25}) satisfies the null condition and, therefore, 
the null Lagrangians for all the considered population dynamics systems 
are of the form
\begin{equation}
L_{null,i}(\dot w_i,w_i)=\dot w_i e^{I_{\alpha} (w_i)} + C_o\ .
\label{eq40}
\end{equation}
It is easy to verify that substitution of $L_{null,i}(\dot w_i,w_i)$
into the E--L equation gives identically zero. 

Now, to derive the equation of motion given by Eq. (\ref{eq26}),
the null condition of Eq. (\ref{eq37}) must be modified to 
account for the dissipative force.  Then, the null condition is 
\begin{equation}\
     \frac{d L_{null,i}(\dot w_i,w_i)}{dt} = {\mathbb F}_{dis,i} 
(\dot w_{i}, w_{i}) e^{2I_{\alpha}(w_i)}\ .
\label{eq42}
\end{equation}
Substitution of $L_{null, i}(\dot w_i,w_i)$ into this equation gives 
the required equation of motion (see Eq. \ref{eq26}).  In the 
following, we use Eq. (\ref{eq40}) to find the NLs for all the
population dynamics models considered in this paper.

Another important characteristic of NLs is the fact that they 
can always be expressed as the total derivative of a scalar
function [39], which has been called a gauge function [45-49].
Thus, we may write 
\begin{equation}
L_{null,i}(\dot w_i,w_i) = \frac{d \Phi_i(w_i, t)}{dt}=
\frac{\partial \Phi_i}{\partial t} + \dot w_i \frac{\partial 
\Phi_i}{\partial w_i}\ ,
\label{eq43}
\end{equation}
and use it to determine the gauge functions $\Phi_i(w_i, t)$
for all derived null Lagrangian $L_{null,i} (\dot w_i,w_i)$.  

The results of our derivations of the null Lagrangians (see Eq. 
\ref{eq40}) and their corresponding gauge functions (see Eq. 
\ref{eq43}) are presented in Table 2. To the best of our knowledge, 
these are the first null Lagrangians and gauge functions ever 
presented for any biological systems, especially for the 
population dynamics models.\\
%
%%%%%%%%%%%%        Table 2       %%%%%%%%%%%%%%%%%
\begin{table}
\begin{center}
\caption{Null Lagrangians and their gauge functions for the population models}\label{tab2}%
\begin{tabular}{@{}lll@{}} 
\hline
{\bf Population models}	&{\bf Null Lagrangians } &{\bf Gauge functions}\\ 
\hline
Lotka-Volterra  &	$L_{null,1} = \dot{w_1} w_1^{-1} + C_0$ 
                &   $\Phi_{1} = \ln \vert w_1 \vert + C_0 t$ \\
			&	$L_{null,2} = \dot{w_2} w_2^{-1} + C_0$ 
                &   $\Phi_{2} = \ln \vert w_2 \vert + C_0 t$\\ \hline
            
Verhulst  	    &	$L_{null,1} = \dot w_1 w_1 ^ {-(1+b)} + C_0$ 
                &   $\Phi_{1} = - b^{-1} w_1^{-b} + C_0 t$ \\
			&	$L_{null,2} = \dot w_2 w_2 ^ {-(1+B)} + C_0$ 
                &   $\Phi_{2} = - B^{-1} w_1^{-B} + C_0 t$\\ \hline

Gompertz  	    & 	$L_{null,1} = \dot{w_1} w_1^{-1} + C_0$ 
                &   $\Phi_{1} = \ln \vert w_1 \vert + C_0 t$ \\
			&	$L_{null,2} = \dot{w_2} w_2^{-1} + C_0$ 
                &   $\Phi_{2} = \ln \vert w_2 \vert + C_0 t$\\ \hline
   
Host-Parasite	&	$L_{null,1} = \dot w_1  w_1^{-1}e^{B \over w1} + C_0$ 
                &   $\Phi_{1} = - Ei ({B \over w_1}) + C_0 t$ \\
			&	$L_{null,2} = \dot{w_2} w_2^{-2} + C_0$ 
                &   $\Phi_{2} = - w_2^{-1} + C_0 t$\\ \hline
                
SIR  		    &	$L_{null,1} = \dot{w_1} w_1^{-1} + C_0$ 
                &   $\Phi_{1} = \ln \vert w_1 \vert + C_0 t$ \\
	       	&	$L_{null,2} = \dot{w_2} w_2^{-1} + C_0$ 
                &   $\Phi_{2} = \ln \vert w_2 \vert + C_0 $t\\ \hline
\end{tabular}
\end{center}
\end{table}
%%%%%%%%%%%%%%%%%%%%%%%%%%%%%%%%%%%%%

\section{Discussion and comparison to the previous work}

Different methods to construct standard and nonstandard Lagrangians 
have been developed and applied to various dynamical systems.  In the 
constructed standard Lagrangians, the presence of the kinetic and 
potential energy-like terms is evident [1-15,32].  However, the 
previously constructed NSLs have different forms, in which such 
energy-like terms cannot be easily recognized [12-31]. Some 
previously obtained NSLs were of the forms given by either Eq. 
(\ref{eq1a}) or Eq. (\ref{eq2a}) [12-15,21,26], but other NSLs 
had significantly different forms [17-19,22-25]. It is important 
to note that the NSLs obtained using the Jacobi Last Multiplier 
method developed by Nucci and Leach (e.g., [17-19]), and those
obtained by El-Nabulsi (e.g., [22-24]) using a different method, 
have been applied to many various dynamical systems in applied 
mathematics, physics, and astronomy.

In the first attempts to obtain Lagrangians for selected biological 
systems, the Lagrangians were found by guessing their forms for
selected population dynamics models [27-29]. Later, those Lagrangians
were formally derived by Nucci and Tamizhmani [30], using the
method of Jacobi Last Multiplier; see also [31] for other applications
of this method to biological systems.  Interesting recent work was 
done by Carinena and Fernandez-Nunez [50], who considered systems
of first-order equations and derived the NSLs that are linear, or 
more generally affine, in velocities using the method of Jacobi 
Last Multiplier. Among the applications of their results to 
different dynamical systems, they also included several 
population dynamics models [50].

In our previous work [32], we derived the first standard Lagrangians 
for five population dynamics models and discussed physical and 
biological implications of these Lagrangians for the models. 
In this paper, we derived the NSLs by using a method that 
significantly modified the one previously developed [12,20].
The results obtained in this paper are now compared to those
previously obtained by Nucci and Tamizhmani [30], as they 
are the most relevant to our work. The comparison shows that 
there are some advantages of using the method of Jacobi Last 
Multiplier, such as the method does not require introducing 
forcing functions and gives directly the same Lagrangians as 
those found earlier in [27-29]. However, the advantage of 
the method developed in this paper is that the derived NSLs
are directly related to their null Lagrangians (NLs) and gauge 
functions (GFs).

Having obtained the NSLs for the considered population dynamics
models, we use them to find their corresponding NLs and then 
to derive the gauge functions (see Table 2); to the best of 
our knowledge, these NLs and GFs are the first obtained for 
biological systems, and specifically, for the population 
dynamics models.  An interesting result is that the NLs
and their gauge functions are identical for the Lotka-Volterra,
Gompertz, and SIR models, which are caused by the same forms of 
NSLs for these three models.  It must also be noted that the 
gauge functions are given by the logarithmic functions, which 
is a novel form among the previously obtained gauge functions 
(e.g., [45-49]). In the approach presented in this paper, 
the main differences between the models are shown by the
introduced dissipative functions. 

Now, the NSLs, NLs and GFs for the Verhulst and Host-Parasite 
models have significantly different forms when compared to the 
other three models (see Table 2).  Moreover, there are also 
differences in their NSLs and NLs. A new result is the form 
of the gauge function $\Phi_1$ for the Host-Parasite model as 
this function is given by the exponential function $Ei(B/w_1)$,
which also appeared in the standard Lagrangian found for this 
system in [32]; this implies that there exist some relationships 
between standard, non-standard and null Lagrangians, and that the 
gauge function plays an important role in such relationships;
further exploration of such relationships will be done in a 
separate paper.

\section{Conclusions}
A new method to derive non-standard Lagrangians for second-order 
nonlinear differential equations with damping is developed and 
applied to the Lotka-Volterra, Verhulst, Gompertz, Host-Parasite, 
and SIR population dynamics models.  For the considered models,
the method shows some limitations, which are explored and it is 
demonstrated that these limitations do not exist for the models 
whose equations of motion can be converted into linear ones. 

The obtained non-standard Lagrangians are different than those 
previously obtained for the same models [30], and the main 
difference is that the previously used Jacobi Last Multiplier 
method does not require introducing dissipative forcing 
functions, which were defined in this paper. However, the 
advantage of the method developed in this paper is that it 
allows us to use the derived non-standard Lagrangians to 
obtain first null Lagrangians and their gauge functions for
the population dynamics models.  

By following the recent work [48,49], the presented results 
also demonstrate how the derived null Lagrangians and gauge 
functions can be used to obtain the equations of motion for 
the considered population dynamics models. Our approach to 
solving the inverse calculus of variation problem and deriving 
non-standard and null Lagrangians is applied to the models of 
population dynamics. However, the presented results show that the method can be easily 
extended to other biological or physical dynamical systems 
whose equations of motion are known.\\

\noindent
{\bf Acknowledgement:} We thank Rupam Das and Lesley Vestal for 
discussions of null Lagrangians and gauge functions and their 
applications to different dynamical systems.\\


\begin{thebibliography}{}

\bibitem{1} J.L. Lagrange, Analytical Mechanics (Springer, Netherlands, 1997). 
\bibitem{2} H. Goldstein, C.P. Poole, J.L. Safko, Classical Mechanics, 3rd Edition 
                  (Addison-Wesley, San Francisco, CA, 2002).
\bibitem{3} J.V. Jos\'e, E.J. Saletan, Classical Dynamics, A Contemporary Approach,
                  (Cambridge Univ. Press, Cambridge, 2002).
\bibitem{4} Lopuszanski, J., The Inverse Variational Problems in Classical Mechanics
                 (World Scientific, Singapore, 1999).
\bibitem{5} N.A. Daughty, Lagrangian Interactions (Addison-Wesley Publ. 
                  Comp. Inc. Sydney, 1990).
\bibitem{6} V.I. Arnold, Mathematical Methods of Classical Mechanics (Springer, 
                    New York, NY, USA, 1978).
\bibitem{7} P.J. Olver, Applications of Lie Groups to Differential Equations
                  (Springer-Verlag, New York, 1993).
\bibitem{8} H. Helmholtz, J.  Reine Angew Math. 100, 213, 1887.
\bibitem{9} J. Douglas, Trans. Am. Math. Soc. 50, 71--128, 1941.
\bibitem{10} S. A. Hojman, J. Phys. A: Math. Gen. 17, 2399--2412, 1984.
\bibitem{11} S.A. Hojman, J. Phys. A: Math. Gen. 25, L291 -- L297, 1992. 
\bibitem{12} Z.E. Musielak, J. Phys. A Math. Theor.  41, 055205, 2008.
\bibitem{13} J.L. Cie\'sli\'nski and T. Nikiciuk, J. Phys. A Math. Theor. 43, 175205, 2010. 
\bibitem{14} Z.E. Musielak, N. Davachi and M. Rosario-Franco, Mathematics. 8, 379, 2020.
\bibitem{15} Z.E. Musielak, N. Davachi and M. Rosario-Franco, J. Appl. Math. ID 3170130 
                   (11 pages), 2020.
\bibitem{16} A.I. Alekseev and B.A. Arbuzov, Theor. Math. Phys. 59, 372--378, 1984.
\bibitem{17} M.C. Nucci and P.G.L. Leach, J. Math. Phys. 48, 123510, 2007. 
\bibitem{18} M.C. Nucci and P.G.L. Leach, J. Math. Phys. 49, 073517, 2008
\bibitem{19} M.C. Nucci and P.G.L. Leach, Phys. Scripta. 78, 065011, 2008
\bibitem{20} Z.E. Musielak, Chaos, Solitons  Fractals, 42, 2640, 2009.
\bibitem{21} A. Saha and B. Talukdar, Rep. Math. Phys. 73, 299--309, 2014. 
\bibitem{22} R.A. El-Nabulsi, App. Math. Lett. 24, 1647, 2011.
\bibitem{23} R.A. El-Nabulsi, Qual. Theory Dyn. Syst. 12, 273, 2013.
\bibitem{24} R.A. El-Nabulsi, Int. J. Theor. Phys. 56, 1159, 2017.
\bibitem{25} F.E. Udwadia, H. Cho, J. Appl. Mech. 80, 041023, 2013.
\bibitem{26} N. Davachi and Z.E. Musielak, J. Undergrad. Rep. Phys. 29, 100004, 2019.
\bibitem{27} E.H. Kerner, Bulletin of Mathematical Biophysics. 26, 333-349, 1964.
\bibitem{28} S.L. Trubatch and A. Franco, J. Theor. Biology. 48, 299-324, 1974.
\bibitem{29} G.H. Paine, Bulletin of Mathematical Biology. 44, 749-760, 1982.
\bibitem{30} M.C. Nucci and K.M. Tamizhmani, J. Nonlinear Math. Phys. 19,
                    12500021, 2012.
\bibitem{31} M.C. Nucci and G. Sanchini, Symmetry, 7, 1613-1632, 2015.
\bibitem{32} D.T. Pham and Z.E. Musielak, Phys. Scripta. submitted, 2022;
                   arXiv:2203.13138v1 [q-bio-PE] 24 March 2022.
\bibitem{33} A.J. Lotka, Elements of Physical Biology (Baltimore, 1925).
\bibitem{34} V. Volterra, Nature. 18, 1-42, 1926.
\bibitem{35} P.F. Verhulst, Correspondance mathématique et physique, 10, 113–21, 1838.
\bibitem{36} B. Gompertz, Phil Trans Roy Soc. 27, 513–85, 1825.
\bibitem{37} V.P. Collins, R.K. Loeffler, H. Tivey, Am J Roentgenol Radium Ther Nuc Med. 
                    78, 988–1000, 1956.
\bibitem{38} W.O. Kermack and A.G. McKendrick, Proc. Roy. Soc. Lond. A. 115, 700-721, 1927.
\bibitem{39} P.J. Olver, Applications of Lie Groups to Differential Equations (Springer-Verlag, 
                   New York, 1993)
\bibitem{40} P.J. Olver and J. Sivaloganathan, Nonlinearity. 1, 389, 1989.
\bibitem{41} M. Crampin and D.J. Saunders, Diff. Geom. and its Appl. 22, 131, 2005. 
\bibitem{42} R. Vitolo, Diff. Geom. and its Appl. 10, 293, 1999.
\bibitem{43} D. Krupka and J. Musilova, Diff. Geom. and its Appl. 9, 225, 1998.
\bibitem{44} D. Krupka. O. Krupkova, and D. Saunders, Int. J. Geom. Meth. Mod. Phys.
                    7, 631, 2010. 
\bibitem{45} Z. E. Musielak and T. B. Watson, Phys. Let. A. 384, 126642, 2020.
\bibitem{46} Z.E. Musielak and T.B. Watson, Phys. Let. A. 384, 126838, 2020.
\bibitem{47} L.C. Vestal, Z.E. Musielak, Physics. 3, 449, 2021.
\bibitem{48} R. Das and Z.E. Musielak, Phys. Scripta. 97, 125213 (12 pages), 2022. 
\bibitem{49} R. Das and Z.E. Musielak, Phys. Scripta. submitted, 2022; 
                    arXiv:2210.09105v1 [math-ph] 17 Oct 2022.
\bibitem{50} J.F. Carinena and J. Fernandez-Nunez, Symmetry. 14, 2520, 2022.                  
\end{thebibliography}
\end{document}